\documentstyle[draft,aps]{revtex}
\newfont{\boldit}{cmbxti10 at 14 pt}
\begin{document}
\title{Neutron-proton interaction in rare-earth nuclei: Role of tensor force}

\author{A. Covello, A. Gargano, and N. Itaco}

\address{Dipartimento di Scienze Fisiche, Universit\`a di Napoli
Federico II \\
and Istituto Nazionale di Fisica Nucleare \\
 Complesso
Universitario di Monte S. Angelo, Via Cintia, 80126 Napoli, Italy}

\date{\today}

\maketitle

\begin{abstract}
We investigate the role of the tensor force in the description of
doubly odd deformed nuclei within the framework of the particle-rotor
model. We study the rare-earth nuclei $^{174}$Lu,
$^{180}$Ta, $^{182}$Ta, and $^{188}$Re using a finite-range
interaction, with and without tensor terms. Attention is focused on  
the lowest $K=0$ and $K=1$ bands, where the effects of the
residual neutron-proton interaction are particularly evident.
Comparison of the calculated results with experimental data evidences
the importance of the tensor-force effects. 

\noindent
PACS number(s): 21.60.Ev, 27.70.+q 
\end{abstract}

\section{Introduction}
It has long been known that
spectroscopy of doubly odd deformed nuclei evidences phenomena which can
be directly associated with the interaction between the unpaired neutron and proton. 
The two most important effects of this kind are known  as  
Gallagher-Moszkowski (GM) splitting \cite{Gallagher58} and Newby (N) shift
\cite{Newby62}.

Recently, use of large multi-detector $\gamma$-ray arrays 
with high analyzing power has largely extended  the amount of experimental
data on this class of nuclei. For
instance, currently
available data in the rare-earth region allow the empirical
determination of 137 GM splittings and 36 N shifts for 25 doubly odd
nuclei \cite{jain98,Headly98}. These are to be compared with 
the 50 GM splittings
and 19 N shifts known in the mid 
1970s, as reported in the extensive review article by Boisson {\it et al.} \cite{Boisson76}.
The new experimental data have also evidenced the occurrence of
other phenomena which may provide further information on the
neutron-proton (n-p) interaction. Of particular interest is the odd-even 
staggering in $K\ne0$ bands, the most important mechanism responsible
for it being the direct Coriolis coupling with one or more N-shifted
$K=0$ bands \cite{jain89}. Also to be mentioned is the signature inversion 
phenomenon, which in the last few years has attracted much attention 
leading to several  experimental and theoretical studies. 
Various attempts have been made to understand this phenomenon 
in term of different mechanisms, but its interpretation still remains
an open question \cite{Garcia01}. 
  
In a previous paper \cite{Covello97} (hereafter referred to as I) we have
studied the doubly odd deformed nucleus
$^{176}$Lu by performing a complete Coriolis band-mixing calculation within the 
framework of the particle-rotor model. We focused attention on
the lowest $K^{\pi}=0^-$ and  $K^{\pi}=1^+$ bands, where the effects of the residual
n-p interaction are particularly evident producing a significant N shift in 
the former and an odd-even staggering in the latter. 
The aim of our study was in fact to investigate the role of the tensor force,
which has long been  a controversial matter.
A detailed discussion on this point can be found in I.  
In that paper we have used for the n-p interaction
both a finite-range force with a Gaussian radial shape and a
zero-range interaction. To completely explore the
role of the tensor force, we have performed two different calculations
with the Gaussian potential, with and without the tensor terms,
respectively.
The main result of these calculations was that the tensor force 
is absolutely essential for 
the description of the above mentioned bands in $^{176}$Lu. In I we have also 
confirmed the relevance  of exchange forces, which was already emphasized in 
Refs. \cite{Frisk88,Nosek94}.

Here, we extend our calculations to other doubly odd deformed
nuclei in the rare-earth region, focusing attention on  N-shifted and
odd-even staggered bands in $^{174}$Lu,
$^{180}$Ta, $^{182}$Ta, and $^{188}$Re (some preliminary results have already
been reported in Refs. \cite{itaco99,itaco01,itaco01a}).
The main motivation for this work  is to verify if the  conclusions 
based on the study of
$^{176}$Lu are confirmed when considering a larger set of experimental data.
In fact, the role of the tensor force in doubly
odd deformed nuclei seems not yet
completely recognized, as evidenced by the large number  of recent calculations
making use of zero-range forces (see for example Ref. \cite{Reviol99} 
and references therein).

The outline of the paper is as follows. In Sec. II we give a brief
description of the model and some details of our calculations.
Our results are presented and compared with the experimental data in
Sec. III. In Sec. IV we draw the conclusions of our study.

\section{Outline of the Model and Calculations}
As mentioned in the Introduction, our calculations are performed within the
framework of the particle-rotor model, in which 
the unpaired neutron and proton are strongly coupled
to an axially symmetric core and interact through a residual effective 
interaction. This is a well known model and a detailed description 
can be found, for instance, in Refs. \cite{Boisson76,Davidson68}. 
For the sake of
completeness, we give in the following the most relevant formulas.     
The total Hamiltonian is written as
\begin{equation}
H=H_0 + H_{{\rm RPC}} + H_{{\rm ppc}} + V_{np}.
\end{equation}
The term $H_0$ includes the rotational energy of the whole system, 
the deformed, axially 
symmetric field for the neutron and proton, and the one-body intrinsic contribution
from the rotational degrees of freedom.
It reads
\begin{equation}
H_0 = \frac{\hbar^2}{2 {\cal J}} ( {\bf I}^2 - I_3^2 ) + H_n + H_p +
\frac{\hbar^2}{2 {\cal J}} [ ( {\bf j}_n^2 - j_{n3}^2 ) + 
( {\bf j}_p^2 - j_{p3}^2 ) ].
\end{equation}
The two terms $H_{{\rm RPC}}$ and $H_{{\rm ppc}}$ in Eq. (1) stand for 
the Coriolis coupling
and the coupling of particle degrees of freedom through the rotational motion,
respectively. Their explicit expressions are
\begin{equation}
H_{{\rm RPC}} = - \frac{\hbar^2}{2 {\cal J}} (I^+J^- + I^-J^+),
\end{equation}
and
\begin{equation}
H_{{\rm ppc}} = \frac{\hbar^2}{2 {\cal J}} (j_n^+j_p^- + j_n^-j_p^+).
\end{equation}
In Eqs. (2)-(4), $\cal J$ is the moment of inertia of the core while $\bf I$
and ${\bf J}={\bf j}_{p} + {\bf j}_{n}$ are the total and intrinsic angular
momentum operators, respectively. $I_3$ and $J_3$ are their projections on 
the intrinsic symmetry axis and, owing to the axial symmetry, are 
represented by the same quantum number $K$.  

The effective n-p interaction in (1) has the general form
\begin{equation}
V_{np}=V(r) [ u_0 + u_1 \mbox{\boldmath {$\sigma$}}_p \cdot \mbox{\boldmath{$\sigma$}}_n + 
u_2 P_M +u_3 P_M \mbox{\boldmath {$\sigma$}}_p \cdot \mbox{\boldmath {$\sigma$}}_n 
+ V_T S_{12} + V_{TM} P_M S_{12} ],
\end{equation}
where the notation is just the same as that adopted in Ref. \cite{Boisson76}.

As basis states we use the eigenfunctions 
of $H_0$, which  
are written as a symmetrized  product  of rotational functions  $D^I_{MK}$
and intrinsic wave functions 
$| \nu _n \Omega_n \rangle | \nu_p \Omega_p \rangle$
\[
| \nu _n \Omega_n \nu_p \Omega_p I M K \rangle =  
\left ( \frac{2I + 1}{16 \pi^2} 
\right )^{\frac{1}{2}} [ D^I_{MK} | \nu _n \Omega_n \rangle | \nu_p \Omega_p
\rangle 
\]
\begin{equation}
~~~~~~~~~~~~~~~~~~~~~~~~~~ + (-)^{I+K} D^I_{M-K}  
|\nu_n \overline{\Omega}_n \rangle | \nu_p \overline{\Omega}_p \rangle ]. 
\end{equation}
Here $\Omega$ is the the quantum number corresponding to $j_3$ 
and $\nu$ stands for all the additional quantum numbers necessary to completely
specify the states. The state 
$| \nu \overline{\Omega} \rangle$ is the time-reversal
partner of $| \nu \Omega \rangle$.  The quantum number $K$ 
has two possible values 
\begin{equation}
K_{\pm}=||\Omega_{n}| \pm |\Omega_{p}||, 
\end{equation}
corresponding to parallel or anti-parallel coupling of 
$\Omega_n$ and $\Omega_p$.   

Each intrinsic state gives then rise to a rotational band, but 
the Hamiltonian (1) 
produces  an admixture of different bands. 
The explicit expressions of the matrix elements of the total 
Hamiltonian can be found in Ref. \cite{Davidson68}, 
where it can be seen that while
the Coriolis interaction has only $\Delta K= \pm 1$ matrix elements,
the two terms $V_{np}$ and $H_{\rm ppc}$ give rise to diagonal and
non-diagonal contributions. In particular, the latter has diagonal matrix 
elements different from zero only for $K=0$ bands with $|\Omega_n|=
|\Omega_p|=1/2$. We would like to stress that all diagonal and 
non-diagonal terms of Hamiltonian (1) are explicitly taken into account in our 
calculations.  

As regards the single-particle Hamiltonians $H_n$ and $H_p$, 
they have been generated by a standard Nilsson potential
as defined, for instance, in \cite{Gustafson67}.
In our calculations  for the four nuclei $^{174}$Lu,  $^{180}$Ta, 
$^{182}$Ta, and $^{188}$Re
the parameters $\mu$ and $\kappa $ in this potential  have been fixed 
by using the mass-dependent
formulas of Ref. \cite{Nilsson69}, while the harmonic oscillator 
parameter $\nu$
has been chosen according to the expression  $\nu= m \omega/ \hbar$ 
fm$^{-2}$. The deformation
parameter $\beta_2$ has been deduced for each doubly odd nucleus 
from the even-even neighbor. In Table I,
we report the values of $\beta_2$ together with those of the rotational 
parameter, $\hbar^{2}/2 {\cal J}$,
used in our calculations. For each doubly odd nucleus this  
latter quantity has been deduced from low-lying bands with a pure 
rotational character.   
The adopted single-particle schemes for the neutron and proton are listed in 
Table II for the four considered nuclei,  each scheme being essentially 
derived from the experimental spectra of the two neighboring odd-mass nuclei.

As regards the residual n-p interaction (5), we have used  
a finite-range force with a radial dependence $V(r)$ of the Gaussian form
\begin{equation}
V(r) = {\rm exp} (-r^2/r_0^2).
\end{equation}

As mentioned in the Introduction, the main aim of this study is to assess the 
role of the tensor force. To this end, we have performed two different calculations
with the Gaussian potential, with and without the tensor 
terms, respectively. 
In both cases, for all the parameters of the force 
we have adopted the values determined by Boisson {\em et al.} \cite{Boisson76},  the 
only exception being $u_{0}$, which does not contribute either to the GM splttings or to
the N shifts. 
These values, as well as that of $u_{0}$,  are reported in I,
where the reasons of our choice are also discussed.

Before closing this Section it is worth mentioning that we have also
performed calculations making use of a $\delta$ interaction with
a strength of the spin-spin parameter varied over a large range of values. 
We have found that no value of this strength gives a satisfactory description of the N shifts 
and odd-even staggerings in any of the  considered nuclei, thus confirming our
results  for $^{176}$Lu (see I).  
As a consequence, here we will not
compare with experiment the results obtained by using a $\delta$ 
force, but make only some comments to point out the inadequacy
of this force to explain the above mentioned effects.

\section{Results and Comparison with Experiment}
We report here the results of our calculations for $^{174}$Lu, $^{180}$Ta, 
$^{182}$Ta, and $^{188}$Re, which are all well deformed nuclei. 
The choice of these nuclei is motivated by the fact that enough experimental
information relevant to our study is available for them. In fact,
in the experimental spectra of $^{174}$Lu, $^{182}$Ta, and
$^{188}$Re  
$K=0$ bands, which are not strongly perturbed but have  a sizeable N shift,
have been recognized,  while in 
$^{180}$Ta and $^{188}$Re  
there are $K=1$ bands which exhibit a significant odd-even staggering. 
We have focused attention on those data which are directly related to
the n-p interaction, so  as to give a sound answer to the question
regarding the role of the tensor term.
In fact, aside from the residual n-p interaction, also the terms
$H_{\rm ppc}$ and $H_{\rm RPC}$ of the Hamiltonian (1) 
may give a contribution to the odd-even shift (see Ref. \cite{jain89})
and a complex
interplay of all these interactions may occur in  bands characterized by a strong admixture of different intrinsic
wave functions. In these cases, even taking explicitly  into account 
all contributions
coming from Hamiltonian (1), it would be a very difficult task to
disentangle the genuine effects of the n-p interaction.   

In Figs. 1-3 the experimental spectra of the $K=0$ bands \cite{Headly98}
are reported and compared with the calculated ones, which are obtained making
use of the two different interactions mentioned in Sec. II.
The $K^{+}=0^{+}$ $p\frac{7}{2}[404]n\frac{7}{2}[633]$
band of $^{174}$Lu starting at 281 keV excitation energy is shown in Fig. 1.
This band, recognized in several other odd-odd nuclei, has been observed in $^{174}$Lu
up to $I^{\pi}=9^{+}$.
We see  that a very good agreement with experiment is obtained for the Gaussian force with
tensor terms. This is not the case for the calculations performed with the pure central
interaction. From Fig. 1 it appears that the two calculations yield almost the same level
spacings for states with even and odd angular momenta separately. The main difference between the
two spectra resides in the energy displacement of levels of odd $I$ relative to those of even $I$,
i.e. the N shift.
In first order perturbation theory the
N shift is just defined as the matrix element
$\langle \nu_{p} \Omega_{p}, \nu_{n} \overline{\Omega}_{n}|V_{np}
| \nu_{p} \overline{\Omega}_{p}, \nu_{n} \Omega_{n} \rangle$.
In Table III we report the values of this matrix element 
obtained with the central plus tensor force together with
the individual contributions. 
For the above mentioned $K=0$ band in $^{174}$Lu we see that the magnitude of the
matrix element is  almost entirely  determined by the tensor force, 
the central part
giving only 1 keV contribution. This result is in line with the early findings of Refs.
\cite{Boisson76,Jones71}, where it was evidenced  the importance of the 
contribution of tensor force
to the N shift for bands based on antiparallel or parallel intrinsic spins. For the latter
the effects of the tensor part are predominant, as expected from the study 
of Ref. \cite{Newby62}, where an analysis 
in the asymptotic limit was performed. In this context, we should also mention that
when using a $\delta$ force, as defined for instance in I,  
there is no reasonable way to reproduce the odd-even shift of this band.
In fact, taking  a negative  value for the strength $v_{1}$ of the spin-spin 
term, the odd-even shift has the opposite
sign  as compared to the experimental one, its magnitude increasing almost 
linearly with
$v_{1}$. As an example, when using $v_{1}=$-0.9 MeV, which is a
reasonable value to reproduce the GM splitting in the rare earth
region \cite{jain98,Elmore76}, the matrix
element $\langle \nu_{p} \Omega_{p}, \nu_{n} \overline{\Omega}_{n}|V_{np}
| \nu_{p} \overline{\Omega}_{p}, \nu_{n} \Omega_{n} \rangle$ becomes
22 keV. The fact that a $\delta$ force produces a shift with the wrong sign while
a pure central force gives a very small contribution, but with the right sign,
makes evident  the importance of the exchange forces.   

In Figs. 2 and 3 we compare the experimental and calculated spectra for two 
lowest $K=0$ bands in
$^{182}$Ta and  $^{188}$Re, respectively. The first is the
$K^{+}=0^{-}$ $p\frac{7}{2}[404]n\frac{7}{2}[503]$ band with a
bandhead energy at 558 keV while the second is the
$K^{+}=0^{+}$ $p\frac{9}{2}[514]n\frac{9}{2}[505]$ band starting at 208 keV.  
Unfortunately, only low-spin members of these two bands 
have been identified, but, as for $^{174}$Lu, an excellent 
agreement is obtained in both cases when a Gaussian force with tensor terms is used.
From Figs. 2 and 3 we see that the right level order is obtained while the 
discrepancies in the energies are less than 20 keV for all the states.
As regards the calculations with a pure central force, the situation is quite 
similar to that discussed for $^{174}$Lu. In fact, the disagreement between 
the experimental data and the results
of these calculations resides in the relative displacement in the
energy of the states with odd and even angular momentum, which, as mentioned before, is
essentially due to n-p interaction. In Table III we report the values of 
the relevant matrix elements. We see again that the pure central force 
gives only a small contribution, the magnitude of these matrix elements 
being essentially determined by the tensor force.     
 
To conclude this discussion, we have shown that the tensor force is quite
able to explain
the odd-even shift in $K=0$ bands of $^{174}$Lu, $^{182}$Ta and $^{188}$Re. 
We now compare
in Table IV the empirical GM splittings \cite{jain98} for these three nuclei
with those derived from the calculated bandhead
energies.
In this Table we only report the results obtained using the
Gaussian plus tensor force. This is to show how this
force, parametrized as in Ref. \cite{Boisson76}, leads  to a very good
description of the three nuclei $^{174}$Lu, $^{182}$Ta and $^{188}$Re
on the whole. In fact, we see from Table IV that not only is the sign
correctly reproduced in all cases, also the quantitative agreement is quite
satisfactory. The discrepancy between the calculated and experimental values 
is always less than 50 keV, except in one case where
it turns out to be 101 keV. In this case, however, 
the empirical GM splitting is affected by an error 
of 57 keV. 

Let us now come to the odd-even staggering effect. In Fig. 4 the experimental
\cite{Dracoulis98,Saitoh99} ratio $[E(I) -E(I-1)]/2I$  for the 
$K^{\pi} = 1^+ p \frac{7}{2} [404] n \frac{9}{2} [624] $
ground-state band in $^{180}$Ta is plotted vs $I$ and compared with
the calculated ones. We see that this band exhibits a rather large
odd-even staggering, the even $I$ states being energetically favored
with respect to the odd ones. This behavior is
satisfactorily reproduced by the calculation including the tensor force while
with the pure central Gaussian force the staggering is almost
nonexistent. It should be pointed out that the staggering in this band may be
traced to direct Coriolis coupling with the $K^{\pi} = 0^+ p
\frac{7}{2} [404] n \frac{7}{2} [633] $ band. In fact, in both our
calculations we have found that the wave functions of the states of
the $K^{\pi} = 1^+$ band contain significant components (up to 30$\%$)
of states of the above $K^{\pi} = 0^+$ band.
The fact that only the
calculation including the tensor terms gives the right staggering
implies therefore that only this force is able to produce 
a sizeable N shift for this  $K^{\pi} = 0^+$ band.
Since this band has not been
recognized in $^{180}$Ta, a direct comparison is not possible at the
present time. It should be noted, however,  that the 
$K^{\pi} = 0^+$ band in $^{174}$Lu (see Fig. 1) has just to the same
intrinsic n-p configuration. As we have already pointed out above,  
only with the Gaussian plus tensor force the
energy of the even $I$ states of this band is decreased so as to bring
the calculated  spectrum in
good agreement with the experimental one.

In Fig. 5 the experimental staggering observed 
in the $K^{\pi} = 1^- p \frac{5}{2} [402] n
\frac{3}{2} [512] $ ground-state band of $^{188}$Re \cite{Headly98} is 
compared with the
calculated ones. Although this band does
not exhibit a large odd-even staggering, we see that only with the 
Gaussian plus
tensor force the calculated behavior reproduces the experimental one.
Also in this case the wave
functions of the states of the $K^{\pi} = 1^-$ band obtained from
both calculations contain appreciable components
(5 - 10 $\%$) of states with $K^{\pi} = 0^- p \frac{5}{2} [402] n
\frac{5}{2} [503] $.
The difference in the two calculations may therefore be traced back to the
different values of the matrix element $\langle p \frac{5}{2} [402]
n \overline{\frac{5}{2}} [503] | V_{np} | p \overline{\frac{5}{2}} [402]
n \frac{5}{2} [503] \rangle $. In fact, as can be seen from Table III,
only the matrix element of the Gaussian plus tensor force turns out to be
adequate to  produce  a sizeable N shift for the $K^{\pi} = 0^-$ band.
Unfortunately, this band has not been observed
in $^{188}$Re or in other rare-earth nuclei.

\section{Concluding Remarks}

In this paper, we have presented the results of a 
study of the well deformed doubly odd nuclei  $^{174}$Lu, 
 $^{180}$Ta,  $^{182}$Ta,  and $^{188}$Re within the framework of the 
particle-rotor model.  We have performed  complete 
Coriolis band-mixing calculations, focusing attention on the lowest
$K=0$ and $K=1$ bands, where the effects of the neutron-proton interaction are
particularly evident.  As regards this interaction, we have used 
in our calculations a Gaussian 
force with and without tensor terms adopting in both cases the values
of the parameters originally determined by 
Boisson {\it et al.} \cite{Boisson76}. 

This study represents an extension of our previous work on $^{176}$Lu aimed at obtaining
further insight into the role of the tensor force in the description of doubly
odd deformed nuclei. The results presented here, as those reported in I, show that
the tensor force is an essential ingredient, if not unique in some cases, to
explain the odd-even shift in $K=0$ bands.  Also the staggering in $K\ne0$ bands,
whose basic mechanism is the Coriolis mixing with N-shifted $K=0$ bands,
is very well accounted for by our calculations with tensor terms. 

It should be mentioned that in more complex situations, where there is
a strong band mixing, 
other mechanisms may play a significant
role and only the right interplay between them can explain the 
observed effects.  This is, for instance, the case of the so called doubly decoupled bands,
namely of bands based on high-$j$ unique-parity shell-model states. 
Also in this context, however,  the tensor force 
should not be neglected, as is done in most of the existing
studies to date. The use of an incomplete neutron-proton interaction 
may in fact lead to an incorrect evaluation of the weight of 
other mechanisms.

\acknowledgements
\noindent
{This work was supported in part by the Italian Ministero dell'Universit\`a 
e della Ricerca Scientifica e Tecnologica (MURST).
NI thanks the European Social Fund for financial support.}

\newpage 

\begin{figure}
\caption{Experimental and calculated spectra of the lowest $K^{\pi} =
0^+$ band in $^{174}$Lu. The theoretical spectra have been obtained by
using (a) a central plus tensor force with a Gaussian radial shape,
(b) a Gaussian central force.}
\end{figure}

\begin{figure}
\caption{Same as Fig. 1, but for the lowest $K^{\pi} =0^-$ band in $^{182}$Ta.}
\end{figure}

\begin{figure}
\caption{Same as Fig. 1, but for the lowest $K^{\pi} =0^+$ band in $^{188}$Re.}
\end{figure}

\begin{figure}
\caption{Experimental and calculated odd-even staggering of the
$K^{\pi} =1^+$ ground-state band in $^{180}$Ta. Solid circles
correspond to experimental data. The theoretical results are
represented by open circles (Gaussian central plus tensor force), and
squares (Gaussian central force). }
\end{figure}

\begin{figure}
\caption{Same as Fig. 4, but for the $K^{\pi} =1^-$ ground-state  
band in $^{188}$Re.}
\end{figure}

\newpage

\mediumtext
\begin{table}
\setdec 0.00
\caption{Values of the parameters $\beta_2$ and $\frac{\hbar^2}{2 {\cal J}}$ (keV).}

\begin{tabular}{lcc}
Nucleus & $\beta_2 $ & $\frac{\hbar^2}{2 {\cal J}}$  \\ 
\tableline
$^{174}$Lu & 0.331 & 12    \\
$^{180}$Ta & 0.278 & 14    \\
$^{182}$Ta & 0.274 & 14    \\
$^{188}$Re & 0.226 & 18    \\
\end{tabular}
\end{table}


\mediumtext
\begin{table}
\setdec 0.00
\caption{Energies (in keV) of the proton and neutron single-particle states.}

\begin{tabular}{lcccc|lcccc}
& Configuration &  & Configuration &  &  & Configuration &  & 
Configuration &  \\
Nucleus & Proton & $E$ & Neutron & $E$ & Nucleus &
Proton& $E$ & Neutron & $E$ \\
\tableline
$^{174}$Lu & $\frac{7}{2}^+ [404]$ & 0 & $\frac{5}{2}^- [512]$ & 0 &
$^{180}$Ta & $\frac{7}{2}^+ [404]$ & 0 & $\frac{9}{2}^+ [624]$ & 0 \\
 & $\frac{1}{2}^- [541]$ & 208 & $\frac{7}{2}^+ [633]$ & 389 &
 & $\frac{9}{2}^- [514]$ & 20 & $\frac{7}{2}^- [514]$ & 200 \\
 & $\frac{5}{2}^+ [402]$ & 369 & $\frac{7}{2}^- [514]$ & 400 &
 & $\frac{5}{2}^+ [402]$ & 250 & $\frac{1}{2}^- [510]$ & 395 \\
 & $\frac{9}{2}^- [514]$ & 437 & $\frac{1}{2}^- [521]$ & 431 &
 & $\frac{1}{2}^+ [411]$ & 546 & $\frac{5}{2}^- [512]$ & 518 \\
 & $\frac{1}{2}^+ [411]$ & 449 & $\frac{1}{2}^- [510]$ & 1053 &
 & $\frac{1}{2}^- [541]$ & 700 & $\frac{1}{2}^- [521]$ & 647 \\
 & $\frac{3}{2}^- [532]$ & 903 & $\frac{5}{2}^+ [642]$ & 1207 &
 &  &  & $\frac{3}{2}^- [512]$ & 785 \\
 & & & & & &  &  & $\frac{7}{2}^- [503]$ & 812 \\
 & & & & & &  &  & $\frac{7}{2}^+ [633]$ & 1506 \\
 & & & & & &  &  & $\frac{5}{2}^+ [642]$ & 2000 \\
\tableline
 & & & & & & & & & \\
$^{182}$Ta & $\frac{7}{2}^+ [404]$ & 0 & $\frac{1}{2}^- [510]$ & 0 &
$^{188}$Re & $\frac{5}{2}^+ [402]$ & 0 & $\frac{3}{2}^- [512]$ & 0 \\
 & $\frac{9}{2}^- [514]$ & 5 & $\frac{3}{2}^- [512]$ & 220 &
 & $\frac{9}{2}^- [514]$ & 176 & $\frac{1}{2}^- [510]$ & 161 \\
 & $\frac{5}{2}^+ [402]$ & 350 & $\frac{11}{2}^+ [615]$ & 300 &
 & $\frac{1}{2}^+ [411]$ & 625 & $\frac{7}{2}^- [503]$ & 310 \\
 & $\frac{1}{2}^+ [411]$ & 477 & $\frac{9}{2}^+ [624]$ & 550 &
 & $\frac{7}{2}^+ [404]$ & 748 & $\frac{9}{2}^- [505]$ & 340 \\
 & & & $\frac{7}{2}^- [503]$ & 670 &
 & & & $\frac{5}{2}^- [503]$ & 640 \\
 & & & $\frac{7}{2}^- [514]$ & 1100 & & & & &  \\
 & & & & & & & & & \\
\end{tabular}
\end{table} 


\mediumtext
\begin{table}
\setdec 0.00
\caption{Values of the matrix elements $\langle \nu_p \Omega_p , \nu_n \overline{\Omega}_n 
| V_{np} | \nu_p \overline{\Omega}_p , \nu_n \Omega_n \rangle $ (in
keV) obtained using a central plus tensor force with a Gaussian radial
shape. $\langle C \rangle $ and $\langle T \rangle $ represent
the contributions of the central and tensor force, respectively.}

\begin{tabular}{lccccc}
 & \multicolumn{2}{c} {Configuration} & & & \\
Nucleus & Proton & Neutron & $\langle C \rangle $ &
$\langle T \rangle $ & $\langle V_{np} \rangle $ \\
\tableline
$^{174}$Lu & $\frac{7}{2}^+ [404] $ & $ \frac{7}{2}^+ [633] $ 
& -1 & -26 & -27 \\
$^{182}$Ta & $\frac{7}{2}^+ [404] $ & $ \frac{7}{2}^- [503] $ 
& 3 & 19 & 22 \\
$^{188}$Re & $\frac{9}{2}^- [514] $ & $ \frac{9}{2}^- [505] $ 
& -4 & -60 & -64 \\
$^{188}$Re & $\frac{5}{2}^+ [402] $ & $ \frac{5}{2}^- [503] $ 
& 0 & 47 & 47 \\
\end{tabular}
\end{table}

\begin{table}
\setdec 0.00
\caption{Experimental and calculated GM splittings (keV).}

\begin{tabular}{lcccccc}
&\multicolumn {2} {c} {\raisebox{0pt}[13pt][7pt] {Configuration}}& & & &\\
Nucleus & Proton & Neutron & \multicolumn {2} {c} {\raisebox{0pt}[13pt][7pt] 
{$K^{\pi}$}} & Expt. & Calc. \\
\tableline 
$^{174}$Lu & $ \frac{7}{2}[404]$ & $\frac{5}{2}[512]$ & $1^-$ & $6^-$ 
&$-114.9 \pm 1.3$ & -124  \\ 
& $ \frac{7}{2}[404]$ & $\frac{7}{2}[633]$ & $0^+$ & $7^+$ &$-64.2 \pm 10.0$ 
& -82  \\ 
&$ \frac{7}{2}[404]$ & $\frac{1}{2}[521]$ & $3^-$ & $4^-$ &$79.5 \pm 5.0$ 
& 89  \\ 
& $ \frac{1}{2}[541]$ & $\frac{5}{2}[512]$ & $2^+$ & $3^+$ &$ -149.8$ 
& -123  \\ 
& $ \frac{5}{2}[402]$ & $\frac{5}{2}[512]$ & $0^-$ & $5^-$ &$130.2 \pm 2.1$ 
& 155  \\ 
& $ \frac{9}{2}[514]$ & $\frac{5}{2}[512]$ & $2^+$ & $7^+$ &$135.3 \pm 3.1$ 
& 109  \\ 
$^{182}$Ta & $ \frac{7}{2}[404]$ & $\frac{1}{2}[510]$ & $3^-$ & $4^-$ 
&$-92.6 \pm 8.3$ & -118  \\ 
& $ \frac{7}{2}[404]$ & $\frac{3}{2}[512]$ & $2^-$ & $5^-$ &$128.0 \pm 0.7$ 
& 142  \\ 
&$ \frac{7}{2}[404]$ & $\frac{7}{2}[503]$ & $0^-$ & $7^-$ &$-122.0 \pm 3.1$ 
&-143  \\ 
& $ \frac{9}{2}[514]$ & $\frac{1}{2}[510]$ & $4^+$ & $5^+$ &$ 147.6 \pm 0.6$ 
& 105  \\ 
& $ \frac{9}{2}[514]$ & $\frac{3}{2}[512]$ & $3^+$ & $6^+$ &$-111.1 \pm 12.4$ 
&-123  \\ 
& $ \frac{9}{2}[514]$ & $\frac{11}{2}[615]$ & $1^-$ & $10^-$ &$147.7 \pm 0.6$ 
& 122  \\ 
& $ \frac{5}{2}[402]$ & $\frac{1}{2}[510]$ & $2^-$ & $3^-$ &$301.2 $ 
& 296  \\
$^{188}$Re & $ \frac{5}{2}[402]$ & $\frac{3}{2}[512]$ & $1^-$ & $4^-$ 
&$-138.7 \pm 1.4$ & -117  \\ 
& $ \frac{5}{2}[402]$ & $\frac{1}{2}[510]$ & $2^-$ & $3^-$ &$93.5 \pm 1.1$ 
& 100  \\ 
&$ \frac{5}{2}[402]$ & $\frac{7}{2}[503]$ & $1^-$ & $6^-$ &$209.4 \pm 0.2$ 
& 205  \\  
&$ \frac{1}{2}[411]$ & $\frac{3}{2}[512]$ & $1^-$ & $2^-$ &$197.2 \pm 57.3$ 
& 96  \\
\end{tabular}
\label{Table I}
\end{table}

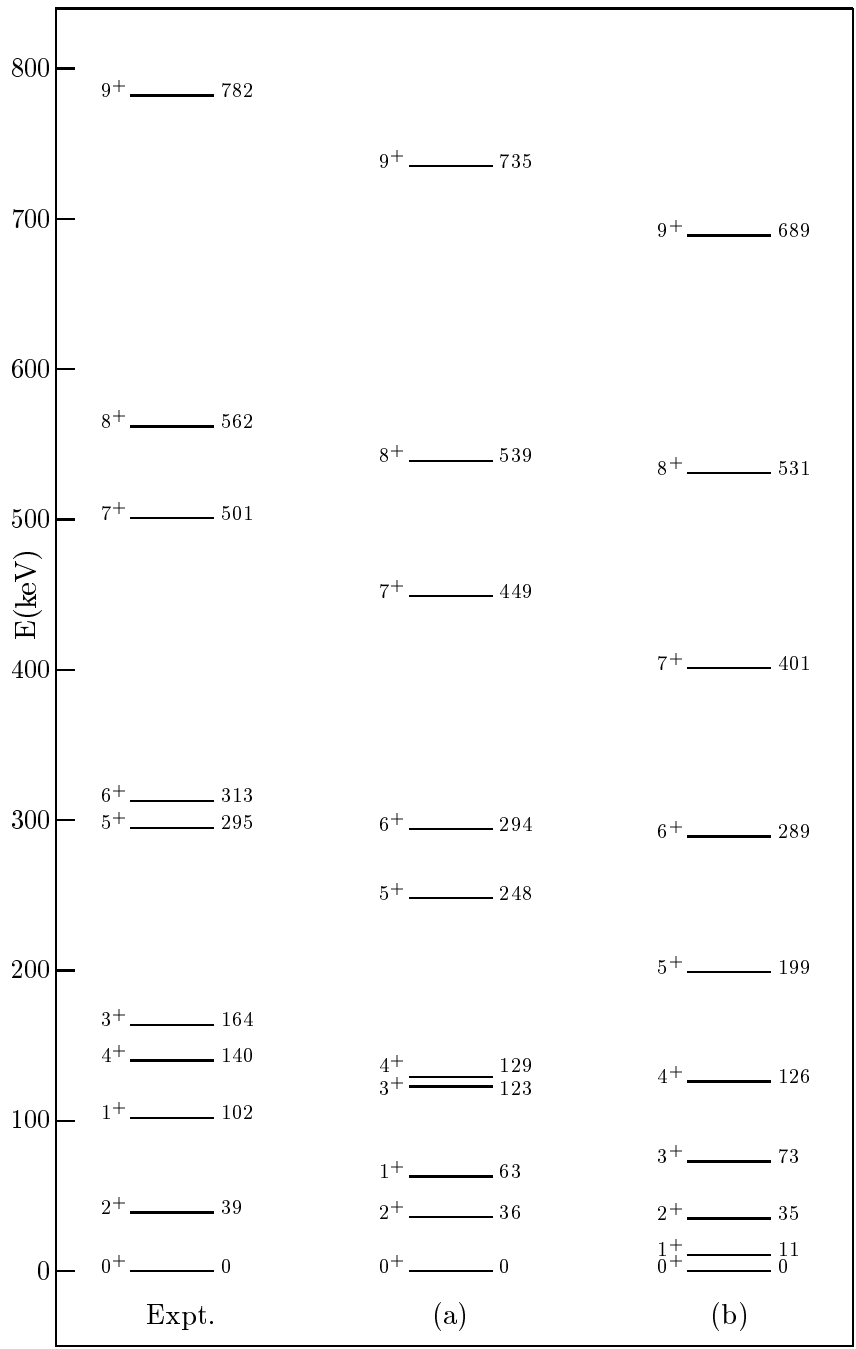
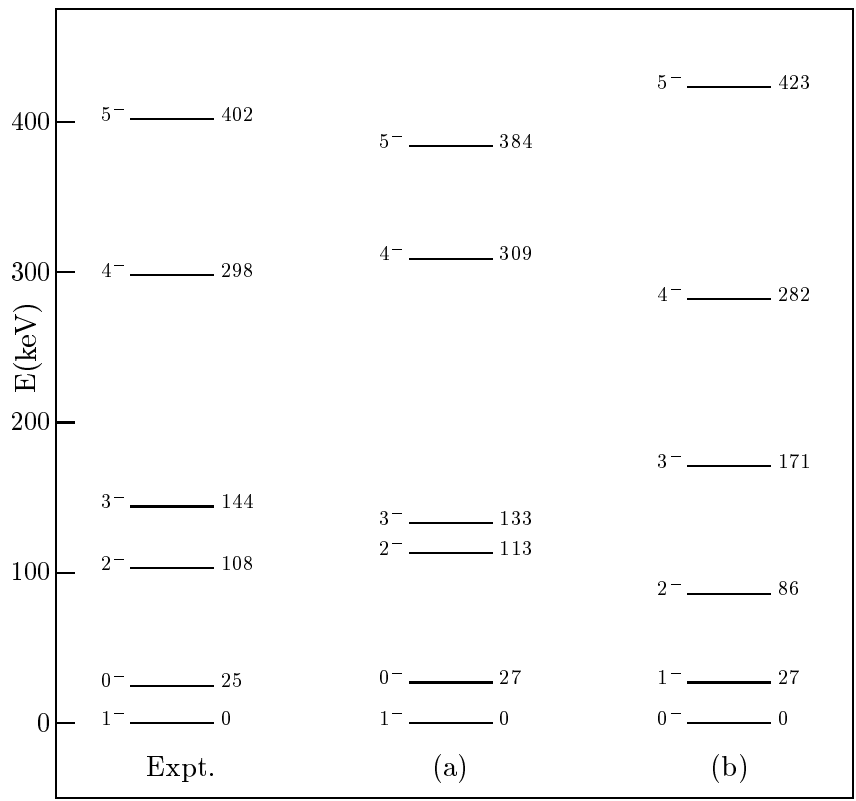
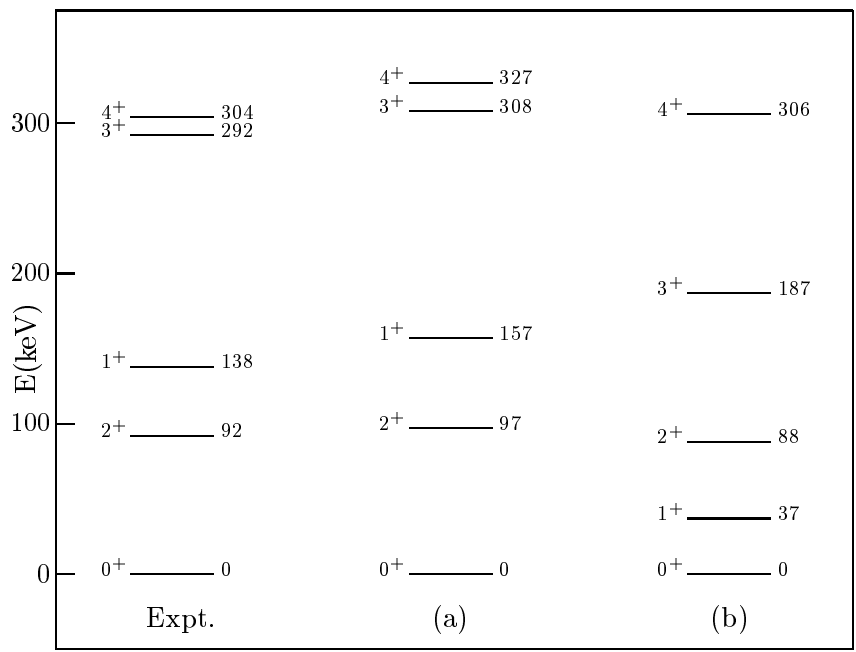
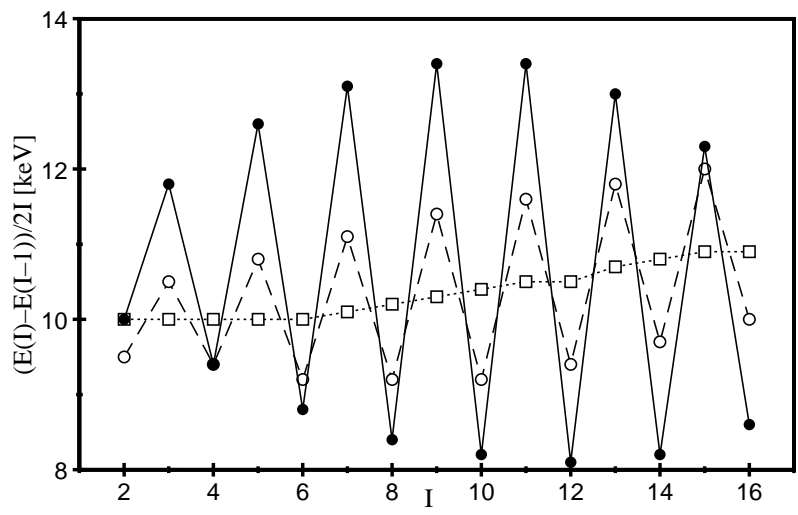
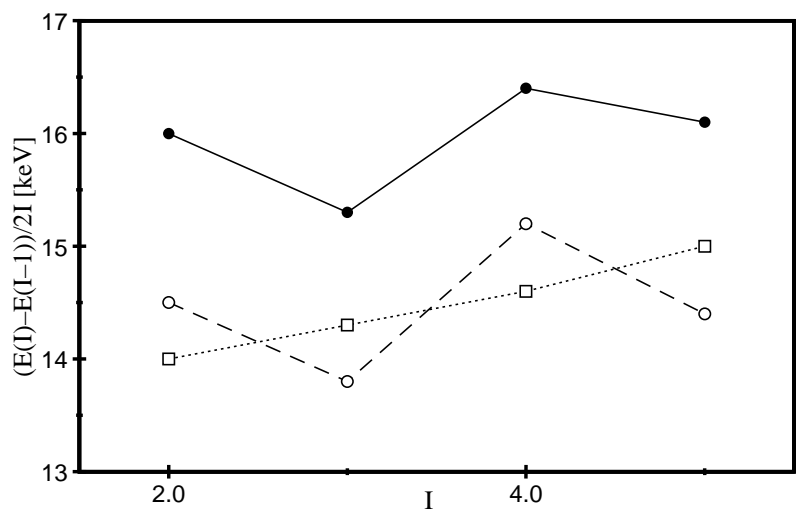
\end{document}